\documentclass[ag]{copernicus}

\begin{document}

\title{The geomagnetic/magneto-telluric induction problem with spatial conductivity fluctuations
}

\author[1]{R. A. Treumann\thanks{Visiting the International Space Science Institute, Bern, Switzerland}}
\author[2]{W. Baumjohann}

\affil[1]{Department of Geophysics and Environmental Sciences, Munich University, Munich, Germany}
\affil[2]{Space Research Institute, Austrian Academy of Sciences, Graz, Austria}

\runningtitle{Geomagnetic Induction Problem}

\runningauthor{R. A. Treumann, W. Baumjohann 
}

\correspondence{R. A.Treumann\\ (rudolf.treumann@geophysik.uni-muenchen.de)}

\received{ }
\revised{ }
\accepted{ }
\published{ }


\firstpage{1}

\maketitle

\begin{abstract}
The electromagnetic induction equation (Helmholtz equation) for the electrically conducting Earth is generalised to the inclusion of a spatially fluctuating internal conductivity spectrum that is superimposed on a one-dimensional large-scale conductivity reference-profile which depends solely on the vertical coordinate $z>0$. This large-scale profile is assumed to be known. It might consist of a stratification of layers of different conductivity and thickness but serves just as the background reference for a distribution of intrusions of differently conducting regions whose size may spatially vary according to a (given or either unknown)  spectrum of fluctuations. The distribution of the fluctuations is allowed to be two-dimensional but can also be three-dimensional. We obtain the conductivity-fluctuation generalised Helmholtz equation. It depends on the spectral density of the fluctuations as function of the vertical and horizontal coordinates. These are contained in an additional term in the Helmhotz equation. In the case of a prescribed fluctuation model the forward problem must be solved. Application to the inverse geomagnetic/magneto-telluric problem in either of its versions is discussed. In its analytical approach \citep{weidelt1972,weidelt2005} we show that the inverse problem yields the fluctuation spectrum. The relation between the spectral density and the obtained conductivity profile is given. By chosing different reference profiles the spectrum could be optimised to fit the data best. Making use of the penetration depth, a relation between the spectral density of the conductivity spectrum is derived which is subject to further geophysical interpretation. This may be of interest in space applications like the JUICE mission.
\end{abstract}

\section{Introduction}
\label{intro}
The present note\footnote{This paper has in 2016 been rejected by Ann.Geophys. with the reasonable comment that the authors are not familiar with the progress achieved within the past 3 decades in solving the induction problem numerically suggesting several reviews \citep{alumbough1996,avdeev2005,egbert2012,kelbert2014,meqbel2014,yang2015}. We are very grateful for these suggestion as we are indeed not working in this field coming from plasma turbulence with what is proposed in this note to possibly be applicable to the induction problem as well. Since our time does not permit to invest more efforts into this problem we simply provide it here for the information of an interested reader. One of the referees also claimed that the fluctuations had been already included in theory. In inspecting the above reviews we have indeed not found any similarity with our approach. Temporal fluctuation spectra have of course been treated but the inclusion of spatial conductivity fluctuations to the degree done here is not detectable to us. There is no doubt that in praxi the numerical work is very valuable. The reviewer wrote: ``In fact, as is shown by dozens of published manuscripts, these reasons include the presence of water and partial melt, temperature variations, presence of metals and certain other volatiles, such as carbon dioxide, in the Earth's mantle and the crust. Moreover, practical 3D inversions are able to constrain some of these features regionally in the Earth." There is nothing to add to this statement. Nevertheless we believe it makes sense, to bring this idea/attempt to the attention of the community without completely reworking.}  deals with an extension of geomagnetic induction theory to the inclusion of a spatially fluctuating conductivity distribution inside the Earth.

In geophysical prospecting, as well as in geomagnetism in general, it is of vital interest to gather information about the conductivity structure of Earth's interior in particular about the distribution of conductivity in Earth's crust and mantle. These regions are not directly accessible such that one is forced to apply indirect methods of determination of the properties of the underneath material. Seismic investigations already provide most important information. However, information about the conductivity requires additional input since the phase velocities of seismic waves depend only indirectly on the underground conductivity. Geomagnetic induction methods are therefore highly desirable. Though being indirect as well, they are bound to the conductivity distribution via the penetration depth of the external geomagnetic field into the ground. This is a function of conductivity and frequency of the applied natural magnetic induction field. Accounting for the induction inside the Earth is important as well for separating the internal contribution to the geomagnetic variation field. An important further application concerns the question of how the internal dynamo-generated geomagnetic field penetrates almost diffusively across the conducting mantle and crust. 

Theoretically, it is quite clear that the most inhomogeneously electrically conducting parts of Earth are the crust and mantle with the scale of inhomogeneity probably increasing with increasing depth below surface. Moreover, below a certain mantle depth and certainly at the transition from mantle to core Earth's interior can, from any practical point of view, be considered as ideally conducting, no matter what the reason for the generation of conductivity will be. This assumption is invalid only when dealing with the geomagnetic dynamo. Since the dynamo field varies on a long scale only, at the corresponding time scales the mantle and crust can be considered as barely conducting. 

Geomagnetic induction theory has found extended application for now decades. In the present note we add to it the proposal that the conductivity structure inside Earth -- and, of course, also on other planets or their satellites like, for instance, Ganymed, if only regular magnetometer measurements of the fluctuations of their magnetic fields will be performed over a longer period on close-by orbits like those planned for the JUICE mission --  can be considered as a superposition of a main and simple one-dimensional given reference-conductivity model, in which the conductivity depends only on depth, and a superimposed fluctuating conductivity of shorter scale which modifies the reference model. We show how the common induction theory has to be modified in order to account for this distribution and can possibly provide additional information about the conductivity structure in the underground up to a certain depth beneath surface. 

\section{Brief review of basics}
The general geomagnetic induction problem is, of course, subject to the quasi-stationary electrodynamics applied to the body of Earth. The idea is to gather information about the conductivity distribution beneath/below Earth's surface down to unaccessible depths by monitoring the Earth's electromagnetic response to variations in the geomagnetic field at the surface. This problem is known as the inverse geomagnetic induction -- or magneto-telluric -- inversion problem. (A similar though different problem is monitoring the variations of Earth's internal field at Earth's surface in view of their information content about the conductivity structure in the layers between the dynamo region and observational sites.) By its nature this is an ill-posed problem. Its assumptions are that the original induction field, a varying magnetic field $\mathbf{B(x,}t)$ -- mostly of natural origin -- induces an electric current flow $\mathbf{j}$ inside the electrically conducting parts/layers of the sub-surface interior of the Earth whose electromagnetic induction field penetrations to the surface where it can be detected as a modification of the initial induction field. The problem is quasi-stationary such that any displacement currents can be neglected. Moreover, there are no electric charges present. Then, with the (isotropic scalar) conductivity $\sigma(\mathbf{x})$, on all relevant time scales a stationary function of location $\mathbf{x}$ only\footnote{The condition of stationarity applies to any of the geomagnetic induction fields considered in the induction problem. However, when being interested in long-scale variations like the secular variation of the geomagnetic field or long-term changes in the external conditions of the geomagnetic field, say secular averages of geomagnetic disturbances etc., then time scales of plate tectonics as well as variations in the geodynamo come into play. Under such conditions the conductivity distribution inside the Earth will exhibit long-term variations and becomes a function of time.}, Maxwell's equations reduce to 
\begin{equation}
\nabla^2\mathbf{E} =\mu_0\sigma(\mathbf{x})\partial_t\mathbf{E}, \qquad \nabla\times\frac{1}{\mu_0\sigma(\mathbf{x})}\nabla\times\mathbf{B} = -\partial_t\mathbf{B}
\end{equation}
with the additional conditions
\begin{equation}
{\nabla}\cdot\mathbf{B} = 0, \qquad \nabla\cdot\mathbf{E}=-\mathbf{E}\cdot\nabla\ln\in(\mathbf{x}) 
\end{equation}
with boundary conditions at Earth's surface requiring that the tangential electric and magnetic fields must be continuous. The latter of the last equations is added here for the reason that the dielectric constant $\vec{\epsilon} =\ \in(\mathbf{x})\cdot\epsilon_0$ may locally deviate from its vacuum value. This is for instance the case in water where one has $\in(\mathrm{H_2O})\approx 80$. Moreover the dielectric constant depends on temperature and frequency. This condition is usually neglected in any treatment of the geomagnetic induction problem. It may become important when interpreting induction effects at the oceanic shore and elsewhere in layers where water is contained. We may assume that $\nabla \in$ is restricted only to the boundaries of such domains, for instance ocean shores. In this case the above condition just becomes a secondary condition on the solution of one of the above field equations, where it must necessarily play a role when considering induction effects of, for instance, the equatorial electrojet or alternatively the geomagnetic Sq-variations. 
  
Since the electric and magnetic fields are related by $\partial_t\mathbf{B}=-\nabla\times\mathbf{E}$, it is most convenient to stay in the electric TE representation, restricting to the electric field. Moreover, Lorentz-gauge implies $\mathbf{E}=-\partial_t \mathbf{A}$, and $\nabla\cdot\mathbf{A}=0$ holding outside boundaries of domains where $\in$ changes discontinuously. Ignoring, as commonly done, such regions, the TE problem can also equivalently be formulated for the vector potential $\mathbf{A}(\mathbf{x},t)$ 
\begin{equation}
\nabla^2\mathbf{A} =\mu_0\sigma(\mathbf{x})\partial_t\mathbf{A}, \qquad \nabla\cdot\mathbf{A}=0
\end{equation}
The magnetic field follows from $\mathbf{B}=\nabla\times\mathbf{A}$.

In locally homogeneous tangential induction fields $\mathbf{E}= E\mathbf{e}_x$, with $\mathbf{e}_x$ the tangential vector at Earth's surface, only one field component remains. This is the simplest imaginable case. If one, in addition, restricts to a mere vertical $z$-dependence of the conductivity the problem simplifies even more. The TE equation then reduces to
\begin{equation}
E''(z,t)=\mu_0\sigma(z)\partial_tE(z,t)
\end{equation}
where $'$ indicates differentiation with respect to $z$. The time dependence can quite generally be treated by Laplace transformation
of either the fields or the vector potential:
\begin{equation}
\mathbf{E}(\mathbf{x},s)= \int\limits_0^\infty \mathrm{d}t\mathbf{E}(\mathbf{x},t)\mathrm{e}^{-st},~~ 
\mathbf{E}(\mathbf{x},t)= \!\!\!\!\!\!\!\!\!\int\limits_{c-i\infty}^{c+i\infty}\!\!\!\!\frac{\mathrm{d}s}{2\pi i}\mathbf{E}(\mathbf{x},s)\mathrm{e}^{st}
\end{equation}
This is preferable as it allows for application to non-periodic perturbations. The energy densities of the electric and magnetic fields are given by
\begin{eqnarray}
\frac{\epsilon_0}{2}\langle\mathbf{E}^2\rangle&=& \frac{\epsilon_0}{8\pi^2}\int_{c-i\infty}^{c+i\infty} \mathrm{d}s \big|\mathbf{E}(s)\big|^2 \\[1ex]
\frac{1}{2\mu_0}\langle\mathbf{B}^2\rangle&=& \frac{1}{8\pi^2\mu_0}\int_{c-i\infty}^{c+i\infty} \mathrm{d}s\Big|\nabla\times\mathbf{E}(s)/s\Big|^2
\end{eqnarray}
The integrands are the spectral energy densities of the fields averaged over the longest time variation.

The transformed fields are complex quantities. The field equation then reduces to the Helmholtz equation conventionally used in the geomagnetic induction problem
\begin{equation}
\nabla^2\mathbf{E}(\mathbf{x},s) = \mu_0\sigma(\mathbf{x})\Big[s\mathbf{E}(\mathbf{x},s)-\mathbf{E}^0\Big]
\end{equation}
 where $\mathbf{E}^0\equiv \mathbf{E}(\mathbf{x},t=0)$. With $z$ the vertical coordinate one may then, as done in TE/magneto-tellurics, define a response function at Earth's surface $z=0$ by 
\begin{equation}
c(x,y,0,s)=-\frac{E_T(x,y,0,s)}{E'_T(x,y,0,s)}=-\frac{E_T(x,y,0,s)}{sB_T(x,y,0,s)}
\end{equation}
where $E_T(x,y,0),~B_T(x,y,0)$ are the local orthogonal tangential components of the electric and magnetic fields at Earth's surface obtained in the point $(x,y)$. The response function $c(x,y,0,s)$ is a measurable quantity. It is an analytical function whose general properties have been investigated. The last equation can also be written as the product of two Laplace transforms
\begin{equation}
E_T(x,y,0,s)= -c(x,y,0,s)E'_T(x,y,0,s)
\end{equation}
Hence, the function $c(x,y,0,s)$ enters the temporal convolution of the vertical derivative of the tangential electric field in any given surface point $(x,y)$:
\begin{equation}
E_T(0,s) = -\int_0^\infty \mathrm{d}t\ \mathrm{e}^{-st} \int_0^t c(0,t-\tau)E'_T(0,\tau)\mathrm{d}\tau
\end{equation}
The response function $c(x,y,0,s)$ folds the field and its vertical derivative. It mediates the time-delayed response to the tangential induction field in this particular surface point. 

Supplied with the boundary conditions of continuity of the tangential fields at Earth's surface, the above TE equation for the electric field and the response function form the basis for the electromagnetic induction problem. Here we have chosen the TE representation. Alternatively an equivalent magnetic formulation can be given which, however, is formally more involved.  

There are several approaches to solve the TE problem. The forward approach assumes a model of the conductivity distribution in order to reproduce and match the magnetic and electric field components which result from the response of the Earth to the application of the external inducing field. The complementary approach is to solve the so-called inverse induction problem. For a large number of measurements of the electric field components or either the  response function the system becomes overdetermined. This enables the application of well-developed numerical procedures to infer about the conductivity structure in one or more dimensions and application of optimisation methods \citep[cf., e.g.,][for application of optimisation methods in the inverse problem]{beusekom2010}. The third more theoretically oriented method relies on the solution of the theoretical inverse problem. This method has been put forward by \citet{weidelt1972,weidelt2005} and is based on the reformulation of the Gel$'$fand-Levitan approach \citep{gelfand1951,marchenko2011} to the inverse Sturm-Liouville (stationary Schr\"odinger equation) problem \citep[cf., e.g.,][for the rigorous mathematical theory of the latter]{koelink2008}. As shown by \citet{parker1980}, the exact solution of this problem in the geomagnetic/magneto-telluric case is possible only in certain cases. Nevertheless, it provides a very lucid understanding of the inversion of the Helmholtz-geomagnetic induction equation and the limitations of any inversions, in particular as the problem itself is ill-posed \citep{jackson1970}. It does anyway not allow for an unambiguous reconstruction of the conductivity profile in the underground. Just from this point of view a statistical approach to the Earth's crustal and mantle conductivity distribution is of vital interest in practical applications as well as it provides a base for further geophysical interpretation.

In the following we propose a reformulation of the above sketched theory to the inclusion of a distribution of conductivity in the Earth. Such an approach circumvents the necessity of solving  a particular given model which might not be well suited in real situations.

\section{Spatially fluctuating conductivity distribution}
Let us assume that the conductivity of Earth has a particular structure which sonsists of two main components. The large-scale structure of the conductivity, $\sigma(z)$ is assumed to be vertical (or radial) with the conductivity organized in shells or otherwise varies gradually on the vertical scale $z>0$ only. This main vertical conductivity profile is overlaid with irregularly distributed smaller regions of lesser or better conducting material. Of such regions there may be many. Treating them in any forward model becomes impossible just for the reason that it would be very difficult to adjust for the geometries and satisfy the various boundary conditions. In order to circumvent this difficulty the scale of variation of the conductivity distribution is assumed to be less than the scale of the primary induction field, in which case the variation can be considered smooth and no internal boundary conditions have to be taken into account. With these assumptions the conductivity is represented as the sum of two terms
\begin{equation}
\sigma(\mathbf{x})= \sigma_0\big[F_0(z) +\delta F(z,\mathbf{x'})\big]
\end{equation}
where $\sigma_0$ is a constant reference value, $F_0(z)$ the main vertical structure function, and $\delta F(z,\mathbf{x'})$ is the  fluctuation of the conductivity, in general a three-dimensional function of the primed small-scale coordinates $\bf{x'}$ that varies slowly with depth on the large scale $z$. One may assume that the smallest scales should be found in the Earth's crust closer to the surface while the scale of the fluctuations should grow with increasing depth. However, at the present stage no such explicit assumptions will be made on the particular form of $\delta F$. The only assumption  is that the distribution of the intrusions of better or worse conducting material is such that on the vertical scale $\delta F$ averages out when averaged over the main reference frame of the main global vertical conductivity profile. 

This assumption does not imply that the fluctuations in conductivity are small amplitude; it merely says that their \emph{scales} on the vertical are substantially shorter than the scale of variation of the main stratification of the conductivity. Nothing is assumed in the horizontal direction except the noted smallness of the scale with respect to the horizontal scale of the applied induction field. Thus the condition is that 
\begin{equation}
\big\langle \delta F\big\rangle = 0
\end{equation}
holds locally, i.e. on the noted horizontal scale. Monitoring the sub-surface response to the application of the temporal variation of the induction field along the surface thus provides information about the lateral variation of the conductivity distribution. The angular brackets $\langle\dots\rangle$ indicate averaging over the vertical scale $z$ of the variation of the main profile $F_0(z)$. This also implies that the fluctuating part of the induction field  and its derivatives that are produced on the short scales by the spatial fluctuations of the conductivity average out: $\langle \delta\mathbf{E}\rangle = \langle \delta\mathbf{E'}\rangle = 0$.

With these conditions in mind we rewrite the basic TE-Helmholtz equation for the above conductivity profile with $\mathbf{E}\to \mathbf{(E}_0+\delta\mathbf{E})$ obtaining, after averaging
\begin{equation}\label{eq-flucavdind}
\nabla^2\mathbf{E}_0 =\mu_0\sigma_0\Big[F_0\left(s\mathbf{E}_0-\mathbf{E}^0\right)+s\big\langle\delta F\delta\mathbf{E}\big\rangle\Big]
\end{equation}
This equation replaces the original TE equation. It contains the initial value of the electric field $\mathbf{E}^0\equiv\mathbf{E}(t=0)$  which results from the use of the Laplace transform. For harmonic time variations of frequency $\omega<\omega_\mathit{max}$ it is more convenient to use the temporal Fourier transform. In this case the initial value drops out, and one has $s=i\omega$. As a new additional term, the above equation contains the correlation between the spatial fluctuations of the conductivity and the electric field that is caused by these fluctuations. 

Subtracting the averaged equation from the original not averaged one, the following Laplace-transformed equation is obtained:
\begin{eqnarray}
\nabla^2\delta\mathbf{E}&=&\mu_0\sigma_0 \Big[s\left(F_0\delta\mathbf{E}+\delta F\delta\mathbf{E}-\big\langle\delta F\delta\mathbf{E}\big\rangle\right) + \nonumber\\
&&+\delta F\left(s\mathbf{E}_0-\mathbf{E}^0\right)\Big]
\end{eqnarray}
It is understood as an equation that determines the spatial evolution of the fluctuations in the applied electric field including its response which results in the variation $\delta\mathbf{E}$ of the electric field. Here the fluctuations in the conductivity are assumed to be given. Since the long-scale variation of the correlation is taken into account in the former equation we assume that the term containing the correlation on the short scales is of second order and can thus be neglected. Moreover, the average term varies on the long scale only. For this reason it is a constant on the scale of the variation of the electric field. We drop it in this approximation though it could in principle be retained  and added to the last term which provides another inhomogeneity. 

With these approximations and conventions we have for the spatial electric field fluctuations which are produced in the presence of a spectrum of fluctuations in the underground conductivity:
\begin{equation}\label{eq-flucind}
\nabla'^2\delta\mathbf{E} = \mu_0\sigma_0\Big[sF_0\delta\mathbf{E} +\delta F\big(s\mathbf{E}_0-\mathbf{E}^0\big)-s\big\langle\delta F\delta\mathbf{E}\big\rangle\Big]
\end{equation}
This is an inhomogeneous linear equation for the electric field fluctuations. After solving it, the result can be used to obtain an expression for the nonlinear term in the fluctuation-modified Helmholtz equation (\ref{eq-flucavdind}). The prime on the $\nabla'$-operator indicates that  in this equation it acts on the short fluctuation scale only. It does not affect the global scale of the average field $\mathbf{E}_0$ nor the dependence of $F_0(z)$. The boundary condition is that the total electric field and its vertical derivative are continuous at $z'=0$, the surface of Earth. It thus applies to the combination of both the average and the fluctuating fields. Since $\sigma=0$ in $z'<0$, the equation for the fluctuations above Earth reduces to the Laplace equation 
\begin{equation}
\nabla'\delta\mathbf{E}=0,\qquad z'<0
\end{equation}

As explained above, we may for simplicity drop the last term on the right in (\ref{eq-flucind}). Then, on the fluctuation scale all indexed coefficients are constant, as is also the initial field as it does not participate in the fluctuations yet. Moreover we do not assume any fluctuation model. The final equation is linear in the fluctuations and can thus be solved by standard methods. We can apply spatial Fourier transforms in $x',y'$ and a Laplace transform in $z'$, counting $z'>0$ positive in the downward direction.  This implies that $\nabla'=i\mathbf{k}$, with $\mathbf{k}= (k_{x'},k_{y'})$ the a two-dimensional wave vector, yielding
\begin{eqnarray}
\Big(k^2-\nabla_z'^2\Big)\delta\mathbf{E}_\mathbf{k}(z')&=&\mu_0\sigma_0\Big[sF_0\delta\mathbf{E}_\mathbf{k}(z')+\nonumber\\
&&+\Big(s\mathbf{E}_0-\mathbf{E}^0\Big)\delta F_\mathbf{k}(z')\Big]
\end{eqnarray}
This equation must now be Laplace-transformed in $z'$, multiplying with $\mathrm{e}^{-\kappa z'}$ and performing the integration over the interval $0\leq z'<\infty$. Its right-hand side provides no problem. One simply replaces all $z'$-dependencies with $\kappa$. The operator on the left yields, however, two new terms. The final result is
\begin{eqnarray}
 \delta\mathbf{E}_\mathbf{k}(\kappa) & =  & \frac{\mu_0\sigma_0\big(s\mathbf{E}_0-\mathbf{E}^0\big)\delta F_\mathbf{k}(\kappa)}{k^2-\kappa^2-\mu_0\sigma_0sF_0} +  \\[1ex]
  & &~~~~~~~~+ \frac{\delta\mathbf{E}'_\mathbf{k}(0)+\kappa\delta\mathbf{E}_\mathbf{k}(0)}{\kappa^2+\mu_0\sigma_0sF_0-k^2}  \nonumber
\end{eqnarray} 
The prime indicates differentiation with respect to $z'>0$, and the argument $0$ indicates that the quantity has to be taken at $z'=0$. Hence this expression includes the boundary values of the spatial electric field fluctuations at the surface of Earth. All these fluctuations depend on the transforms of the spatial fluctuations of the conductivity $\delta F_\mathbf{k}(\kappa)$.

This expression must be used to calculate the average correlation of the field and conductivity fluctuations. This procedure will yield the explicit form of the ultimate average conductivity fluctuation-generalised induction equation (\ref{eq-flucavdind}).

\section{Fluctuation-generalised induction equation}
Our aim is to find the fluctuation-averaged induction equation which tells how the distribution of electrical conductivity on different scales affects the electric induction field. It will contain all the information about the conductivity distribution in the earthly halfspace. In order to achieve this goal we must calculate the average correlation function $\langle\delta F\delta\mathbf{E}\rangle$ between the spatial fluctuations of the conductivity and the induced fluctuations of the electric field.

Calculating this term requires formation of the product of the conductivity and field fluctuations and averaging over the global scales. When representing the fluctuations as the inverse Fourier respectively Laplace transformations of arguments $\mathbf{k,k'}$ and $\kappa,\kappa'$. A number of exponential functions of these arguments arise. The spatial averages over these functions lead to products of Dirac-delta functions of the kind $\delta(\mathbf{k+k'})\delta(\kappa+\kappa')$. Hence, replacing $\mathbf{k'}\to-\mathbf{k},\ \kappa'\to-\kappa$ yields for the correlation function
\begin{eqnarray}
&&\Big\langle\delta F\delta\mathbf{E}\Big\rangle= \frac{1}{(2\pi)^3i}\int\limits_{-\infty}^\infty\mathrm{d}^2k\int\limits_{c-i\infty}^{c+i\infty}\frac{\mathrm{d}\kappa\ }{k^2-\kappa^2-\mu_0\sigma_0sF_0} \times\nonumber\\
&&\qquad\qquad\times\Big\{\mu_0\sigma_0(s\mathbf{E}_0-\mathbf{E}^0)\Big|\delta F_\mathbf{k}(\kappa)\Big|^2 - \\[1ex]
&&\qquad\qquad-\Big[\delta\mathbf{E}'_{-\mathbf{k}}(0)-\kappa\delta\mathbf{E}_{-\mathbf{k}}(0)\Big]\delta F_\mathbf{k}(\kappa)\Big\}\nonumber
\end{eqnarray}
In this expression $|\delta F_\mathbf{k}(\kappa)|^2$ is the spectral density of the conductivity fluctuations. One may note that the field is symmetric in $\mathbf{k}$: it holds that $\mathbf{E}_\mathbf{k}=\mathbf{E}_{-\mathbf{k}}$. In this notation we suppressed the dependence of all quantities on the large scale $z$. 
\begin{figure}[t!]
\hspace{1.5cm}{\includegraphics[width=0.3\textwidth,clip=]{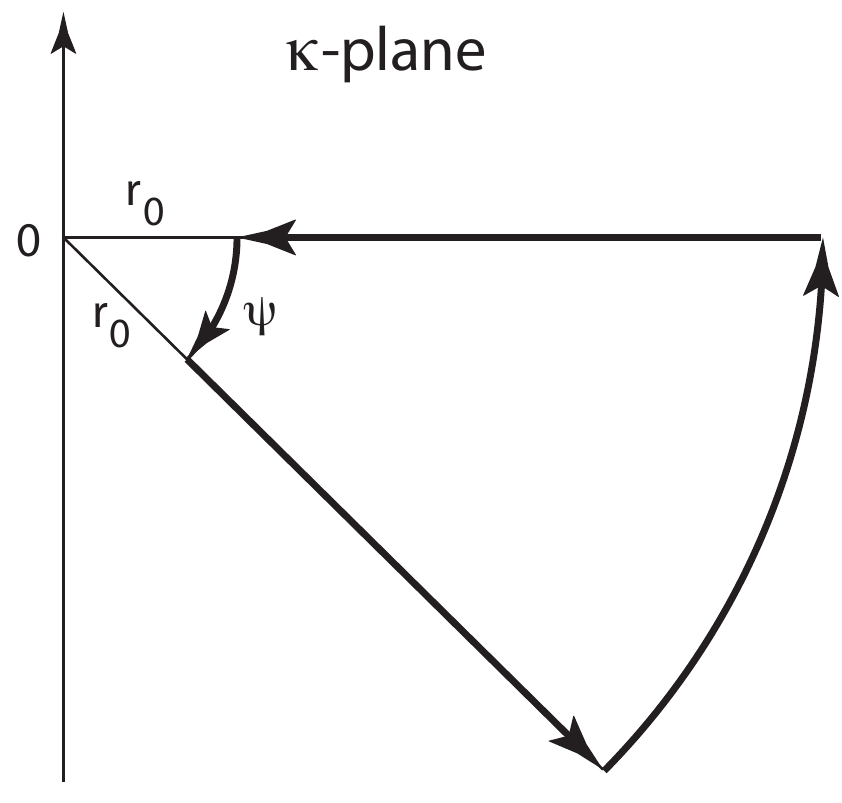} 
}
\caption[]
{The integration contour in the complex $\kappa$ plane. Integration is restricted to the lower $\kappa$-half plane. }\label{fig-int-path}
\end{figure}

In the forward problem one would chose a definite model for the conductivity fluctuations such that it is considered as known. In this version it depends on the large scale coordinates, however.  $F_0({z})$, the main spatial profile of the conductivity, depends on $z$ only. Clearly, this expression is quite involved. It is not only complicated by the presence of the surface and initial values of the field but also by the denominator in the integral. For this reason we will farther below introduce a simplification.

However, before doing this we go one step further in the calculation. The denominator of the complex $\kappa$ integral generates poles in the complex plane. If we assume that the expression in the curly brackets is analytical, then the integral can be solved by integrating over the poles. This can either be done in the $\kappa$-plane or in the $k$-plane. Defining $k_0^2=s\mu_0\sigma_0F_0$ and 
\begin{equation}
\kappa_\pm=\pm\sqrt{k^2-k_0^2} \equiv r\mathrm{e}^{\mp i\psi/2}
\end{equation}
the denominator can be factorised. Under the assumption that the conductivity spectrum behaves regularly, the $\kappa$-integral is formally solved. Both $k$ and $s$ are themselves complex variables. Let us, for simplicity, assume that $s\to i\omega$, with $\omega$ the frequency of the applied induction electric field. It is assumed that a broad range of frequencies may be applied by taking their Fourier transform instead of the Laplace transform. In this case the intial value of the field $\mathbf{E}^0$ drops out from the above equation. We have $k_0^2\equiv i\omega\mu_0\sigma_0F_0=i\rho_0^2$ and complex $k=\rho\,\exp(i\chi)$ with $\rho$ the modulus and $\chi$ the phase. Then
\begin{equation}\label{eq-kappa-k}
\kappa_\pm=\pm\sqrt{\rho^2\exp(2i\chi)-i\rho_0(z)}
\end{equation}
The frequency $\omega$ is real. The poles of $\kappa$ in the complex $(r,\psi)$-plane lie at 
\begin{eqnarray}
r&=&\sqrt[4]{\rho^4\cos^22\chi+\left(\rho^2\sin2\chi-\rho^2_0\right)^2} \\
\psi &=& \tan^{-1}\left(\frac{\rho^2\sin2\chi-\rho^2_0}{\rho^2\cos2\chi}\right)
\end{eqnarray}
In particular $r=\rho_0, \infty$ and $\frac{1}{2}\psi=-\frac{1}{4}\pi, 0$ for $\rho=0, \infty$, respectively. They are conjugate to the positive real axis in this plane. When the complex horizontal wavenumber $k$ changes, they move from $\rho_0\, \to\, \infty$ on two arc sections from $\psi=\mp\frac{1}{4}\pi \, \to\, \psi =0$. Figure \ref{fig-int-path} shows the integration contour in the complex $\kappa$-plane.

After factorisation, the solution of the $\kappa$-integration formally yields the sum of the residua in the $\kappa$-poles
\begin{eqnarray}
&&\Big\langle\delta F\delta\mathbf{E}\Big\rangle=\frac{1}{8\pi^2}\int\limits_{0}^\infty\int\limits_0^{2\pi}\frac{k\,\mathrm{d}k\,\mathrm{d}\phi}{+\sqrt{k^2-k^2_0}} \ 
\Bigg\{\frac{k_0^2}{F_0}\times \\[1ex]
&&\times\Big|\delta F_\mathbf{k}(\kappa_+)\Big|^2\mathbf{E}_0 
-\Big[\delta\mathbf{E}'_{\mathbf{k}}(0)-\kappa_+\delta\mathbf{E}_{\mathbf{k}}(0)\Big]\delta F_\mathbf{k}(\kappa_+)\bigg\}\nonumber
\end{eqnarray}
Of the two residua that with $\kappa=\kappa_+$ is selected by the requirement that the Laplace-transform of $\delta F_k(\kappa)$ exists. We rewrote the two-dimensional volume element of the $\mathbf{k}$-integral in terms of $k$ and the azimuthal angle $\phi$. One may note that the root in the denominator is $\kappa_+$ whose explicit form in terms of $\rho$ and $\chi$ is given above. 

The important term in this expression is the term containing the main electric field $\mathbf{E}_0(z)$. The correlation term adds a linear contribution to the original average induction equation, with the nonlinearity being absorbed into the spectral density of the conductivity fluctuations which appears as a space dependent factor. The appearance of the  boundary values in the last bracket complicates the problem. They represent an inhomogeneity in the fluctuation-averaged induction equation.

Further reduction of the nonlinear correlation term can be done assuming that the fluctuations depend only on $z$ and $x$, a case of practical importance because locally it will always be possible to identify a main direction of strike of the horizontal variation of the conductivity. In this case  the $k$-integration is one-dimensional with $\phi$-integration disappearing. The entire problem is then two-dimensional, and the average  of the conductivity and field fluctuations simplifies to
\begin{eqnarray}\label{eq-average}
\Big\langle\delta F\delta\mathbf{E}\Big\rangle&=&\frac{1}{4\pi}\int\limits_{0}^\infty\frac{\mathrm{d}k\,}{\sqrt{k^2-k_0^2}}\ 
\Bigg\{\frac{k_0^2}{F_0}\Big|\delta F_{k}(\kappa_+)\Big|^2\mathbf{E}_0 - \nonumber\\ 
&-&\Big[\delta\mathbf{E}'_{{k}}(0)-\kappa_+\delta\mathbf{E}_{{k}}(0)\Big]\delta F_{k}(\kappa_+)\Bigg\}
\end{eqnarray}
where $\kappa_+$ is to be expressed through $k$. Solution of the $k$-integral is inhibited if no model is assumed for the conductivity fluctuations. Note again that $\kappa_+, k_0, F_0, \delta F_k$ all depend on the large scale $z$. Hence, the new additional term in the Helmholtz equation caused by the spatial conductivity fluctuations turns out to be a rather complicated function of the global depth-coordinate $z$. 

This expression in its implicit form is to be multiplied with $i\omega$ and inserted into the fluctuation-averaged induction equation (\ref{eq-flucavdind}). While the integral containing the spectral density of the conductivity fluctuations adds to the linear term in that equation, the integral containing the boundary values appears as an inhomogeneity. The TE induction equation thus becomes an \emph{inhomogeneous} Helmholtz equation. 

Because we assumed that the global model is known and is strictly one-dimensional, depending solely on the global vertical conductivity profile $F _0(z)$, the Helmholtz equation remains a one-dimensional equation also when conductivity fluctuations are included. These fluctuations are at least two-dimensional, depending on the vertical and one horizontal coordinate. Their large-scale variation in the  horizontal direction is intrinsic to the $k$-dependence of the new terms. It becomes explicit only after solution of the equation and re-transformation from $k$ to configuration space. The TE problem  could be treated as a forward problem when an appropriate model for the fluctuations of the conductivity is assumed.
Still this is difficult because the coefficients in the Helmholtz equation are not constants. Moreover, its inhomogeneity requires construction of the Green's function. Subsequently the solution becomes an integral of the Green's function and the inhomogeneity.

\begin{figure}[t!]
\hspace{1.5cm}{\includegraphics[width=0.3\textwidth,clip=]{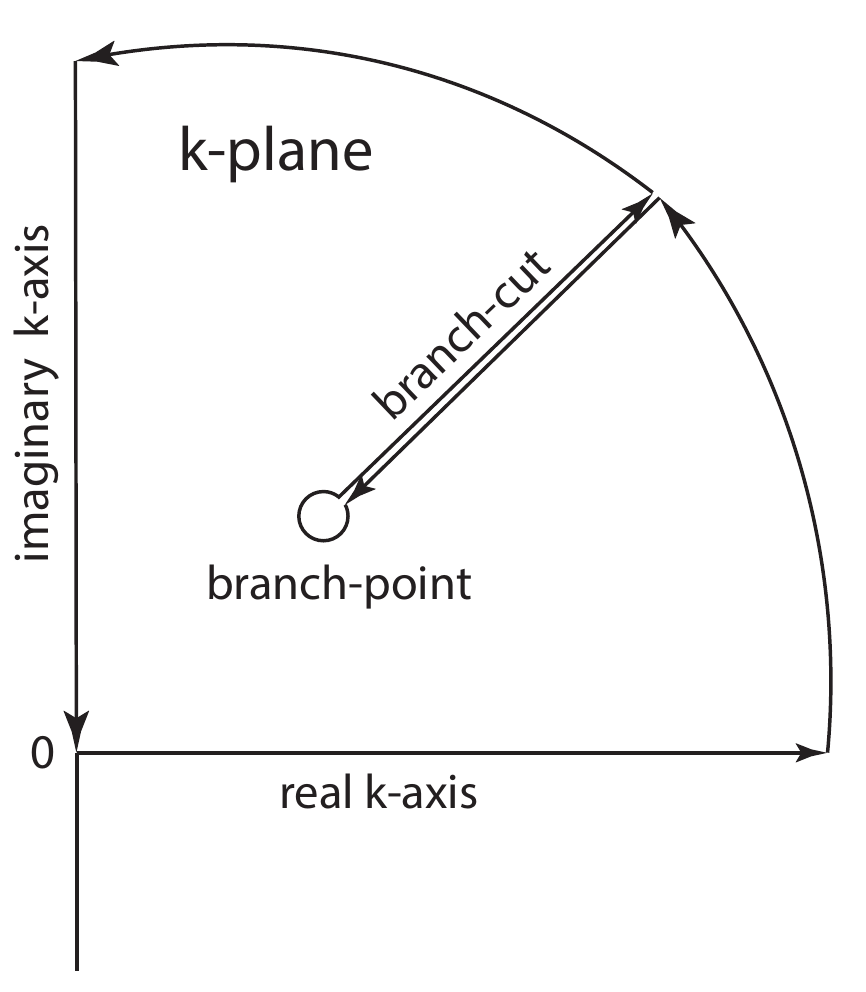} 
}
\caption[]
{The integration contour in the complex $k$ plane. Integration is restricted to the upper right $k$-half plane. The relevant branch point is located  at $k_b=\rho_0\exp(+i\pi/4)$. The branch-cut extends from $k_b$ to infinity where for any reasonable distribution the spectrum vanishes.} \label{fig-k-path}
\end{figure}

The integration variable $k$ is complex. When the residua are entire or analytical functions possessing simple poles in the complex $k$-plane, the presence of the root in the denominator of the integrand introduces two algebraic branch points at 
\begin{equation}
k_{\pm b}=\sqrt{\mu_0\omega\sigma_0F_0}\ \mathrm{e}^{\pm i\pi/4} =\rho_0\ \mathrm{e}^{\pm i\pi/4}
\end{equation}
which depend on $z$ and frequency $\omega$, and one additional branch point at $k = \infty$. The latter provides no further problem because all variables vanish sufficiently fast for $k\to\infty$. Of the two finite branch points only that in the positive $k$-plane is selected for reasons of convergence of the inverse Fourier transform.  The contour of integration in the complex $k$-plane is shown in Fig. \ref{fig-k-path}. Thus, the cut is to be made from the upper branch point $k=k_{+b}\equiv +k_0$ along a line at angle $+\pi/4$ to infinity when integrating with respect to $k$ over the upper $k$ half-plane. In principle, the contour is closed along the imaginary axis from $+i\infty$ down to the origin. It can, however, extended to the entire upper $k$-half plane by analytical continuation. The value of the integral becomes
\begin{eqnarray}
\int\limits_{-\infty}^\infty \frac{\mathrm{d}k}{\kappa_+}\Big\{\dots\Big\}&=& \Bigg[\mathcal{P}\!\!\!\!\!\!\!\!\int\limits_{-\infty}^\infty\frac{\mathrm{d}k}{\sqrt{k^2-k_0^2}} \ \ -2\pi \ \mathrm{Res}(k_{b})\Bigg]\Big\{\dots\Big\}\nonumber \\[1ex]
&=&2\pi i\sum\limits_\mathrm{Res}^{k\neq k_b}\frac{1}{\sqrt{k^2-k_0^2}}\Big\{\dots\Big\}
\end{eqnarray}
It is the sum of the Hilbert-principal value $\mathcal{P}$ of the integral taken along the real $k$-axis, plus the negative residuum in the positive branch point. This sum equals the sum of all residua in the upper positive quadrant of the $k$-plane contributed by the poles of the integrand in the curly braces. All residua contributed by poles on  the real axis must be multiplied by $\frac{1}{2}$. Hence, the $k$-integral is the sum of all residua on the right of the last expression.
If there are no additional poles, i.e. if the spectral density of the conductivity fluctuations is an entire function, then the integral vanishes with vanishing right-hand side. The problem degenerates. The fluctuations in that case do not visibly contribute to the induction field. The above integral, split into imaginary and real parts, then provides a so-called ``dispersion relation'' for the fluctuations. It relates the real and imaginary parts of $k$ to the fluctuation spectrum and the induction electric field. On the other hand, if the $k$-integral in this case is interpreted as a principal-value integral, then the relevant residuum is taken to be the residuum at the branch point: $+2\pi i|\delta F_{k=k_b}(0)|^2$.

The physically important case is when the fluctuation spectrum of the conductivity possesses poles and thus generates non-vanishing residua. This case is realistic because one must expect that the presence of conductivity fluctuations causes a nonnegligible effect in the induced electric field. 

The ultimate fluctuation-generalised Helmholtz equation reads
\begin{eqnarray}\label{eq-avfluct-a}
E''_0(z) &- &k_0^2E_0(z)\Bigg[1+\frac{k_0^2}{2F_0}\sum\limits^{k\neq k_b}_\mathrm{Res}\frac{\big|\delta F_{k}(\sqrt{k^2-k_0^2})\big|^2}{\sqrt{k^2-k_0^2}}\Bigg] \\[1ex]
&=&\frac{k_0^2}{2}\sum\limits^{k\neq k_b}_\mathrm{Res}\frac{\delta F_{k}\left(\sqrt{k^2-k_0^2}\right)}{\sqrt{k^2-k_0^2}} \Big[\delta E'_{k}(0)-\kappa_+\delta E_k(0)\Big]\nonumber
 \end{eqnarray}
where the residua are contributed by the poles of the fluctuation spectrum inside the integration contour in the upper half of the branch-cut complex $k$-plane shown in Fig. \ref{fig-k-path}. It must be stressed that without any assumption about the fluctuations nothing is known about the existence and/or form of the poles in the spectrum of conductivity fluctuations. For this reason the representation given in the above generalised equation is implicit. However, the argument is that any physically reasonable distribution of conductivity fluctuations in the underground, i.e. intrusions of high or low conductivity structures will necessarily give rise to a number of poles or, in more complicated cases, branch points integration around which contribute residua. One might assume that any physical conductivity distribution will not give rise to essential and thus non-integrable singularities. Thus application of the implicit form of the above induction equation is for a large class of fluctuations justified. This class encompasses all fluctuations that satisfy the condition that their scales are shorter than the vertical scale of the reference zero-order model $F_0(z)$. 

The right-hand side of the above euqation contains any possible residua which are contributed by the correlation of the surface fluctuation electric field and the conductivity fluctuations. In a forward problem solution one would determine this term from the boundary conditions. Here, for simplicity, we assume that it can be put to zero because no correlation is expected between the fluctuations in conductivity and field at the surface, a condition which can, in principle, be relaxed when solving the external problem at $z'<0$ for the fluctuations. We restrict to the demonstration of the modification of the induction problem by inclusion of conductivity fluctuations. We therefore don't do this here, relegating the inclusion of the boundary effects to a more extended investigation\footnote{The assumption is that the surface values of the fluctuation-induced electric field variation is of short scale such that the horizontal variation of the electric field is entirely attributed to it. In this case the value of the conductivity-fluctuation-caused field is given by its internal value, and the external field in the non-conducting air just behaves  passively and decays exponentially with increasing altitude above the surface as determined by Laplace's equation.}. Under this simplifying assumption the TE induction equation reduces to zero on the right.

The conductivity fluctuations just generate an additional contribution to the linear term in the Helmholtz equation. Formally this term is  a modification of the conductivity profile while being of second order in the applied frequency. This was expected. Hence, it seems that little has been gained by assuming the presence of a fluctuating conductivity. Moreover, due to the dependence of all quantities on the large-scale coordinate $z$ the spatial variation of the coefficient of $E_0(z)$ is complex. Indeed, a forward analytical solution of the basic fluctuation-averaged TE equation seems improbable even for most simple fluctuation models. Numerical solutions are one way out. 

Let us assume that there are $m$ poles in the upper half $k$-plane which lead to $1\leq n\leq m$ residua with $k_n$ the value of $k$ at the $n$-th pole of the expression under the sum in the above equation. Then, by Fourier-transforming the induction equation back into real configuration space with horizontal coordinate $x$ we obtain its two-dimensional version
\begin{eqnarray}\label{eq-avfluct-b}
&&E''_0(z)\delta(x-x_0)-k_0^2(z)E_0(z)\Bigg[\delta(x-x_0) + \frac{k_0^2(z)}{4\pi F_0(z)}\times\nonumber \\[1ex]
&&\sum\limits_{n=1\atop k_n\neq +k_0}^m\frac{\mathrm{e}^{ik_n(x-x_0)}}{\sqrt{k_n^2-k_0^2(z)}}\bigg|\delta F_{k_n}\left(\sqrt{k_n^2-k_0^2(z)}\right)\bigg|^2\Bigg]=0
\end{eqnarray}
This holds under the strict condition that
\begin{equation}
k_n^2\neq k_0^2= i\rho_0^2
\end{equation}
The field on and along the Earth's surface refers to  $x_0$ as an arbitrary reference point. The horizontal variation of the contribution of the fluctuating conductivity is thus given as a superposition of $m$ harmonic modes with complex wave numbers $k_n$. 

The parameter $\rho_0$ contains the time variation of the induction field. It corresponds to the inverse penetration depth of the electromagnetic field of frequency $\omega$ into the conducting Earth, which thus enters in a rather complicated way. This complication reflects the complex reflection and absorption process of the induction field in the various differently conducting conductivity intrusions. The statistical formulation has the advantage that it avoids the taking into account of many internal conductivity boundaries below Earth's surface in a complicated and still unrealistic structure model of the conductivity. Only the average response of the conductivity structure enters the approach.

\section{Inverse problem}

Solving the above equation in a forward calculation is difficult even when a reasonable conductivity fluctuation model is assumed though it can, of course, be attacked by numerical methods and with the help of supercomputers.

Another way out is the application of the inverse geomagnetic induction problem. The inverse induction problem in either version seeks to gather information about the coefficient of the electric field in the second term of the above equation. It their ordinary version the inverse problems aim on a direct reconstruction of the one- or two-dimensional conductivity $\sigma_0(x,z)$. In the above approach the vertical profile of the conductivity is assumed to be known. Thus, in any given point $x$ on the surface the inversion provides information about the average spectral density of the conductivity fluctuation as function of the global coordinates $z$ and $x$. These coordinates are prescribed by the zero-order model profile. The spectral density thus informs about the average deviation of the real conductivity structure from the model. In principle, this could be used in an optimisation of the vertical model structure by variational methods. In the horizontal direction the restriction is that the distance of measuring points is limited by the scale of the horizontal variation of the external induction field. This is a consequence of the assumption of a homogeneous field.  

In application of the inverse problem it is more convenient to return to the one-dimensional Eq. (\ref{eq-avfluct-a}) than using its $x$-transformed two-dimensional version (\ref{eq-avfluct-b}). The two-dimensionality is secondary only since $k_n$ appears simply as a parameter. With the right-hand side set to zero, the induction equation assumes its canonical form
\begin{equation}
E''_0(z,\omega)=i\omega\mu_0\sigma_0\Sigma(z,\omega)E_0(z,\omega)
\end{equation}
where we understand $E_0(z,\omega) \to E_0(z,\omega)/E_0(0,\omega)$. Moreover, $\sigma_0\Sigma(z)$ is the fluctuation modified equivalent conductivity and
\begin{equation}
\Sigma(z)=F_0\left[1+\frac{k_0^2}{2F_0}\sum\limits_{n=1\atop k_n\neq +k_0}^m\frac{\left|\delta F_{k_n}\left(\sqrt{k_n^2-k_0^2}\right)\right|^2}{\sqrt{k_n^2-k_0^2}}\right]
\end{equation}
where all quantities $\rho_0, F_0, k_n$ are functions of $z$. This dependence complicates the calculation but is, for the inverse problem, not of primary importance. It simply implies that the conductivity structure is not as simple as the originally assumed stratification model that is contained in the function $F_0(z)$. With this normalisation the surface response function at $z=0$ is
\begin{equation}
c(\omega)=-\frac{1}{E'_0(0,\omega)} 
\end{equation}
a function which behaves analytical if only $\Sigma(z,\omega)$ is analytical, since this is one of the conditions which must generally hold for application of our theory. 

In the following we are not primarily interested in a brute force numerical solution of the geomagnetic inversion problem. The reader is referred to a broad literature on this subject \citep[cf., e.g., the reviews by][and references therein]{backus1967,bailey1970,johnson1971,parker1970,parker1980,parker1994,gubbins2004}. Rather we seek to relate it to the analytical formulation of  the one-dimensional inverse problem as given by \citet{weidelt1972}, who first transformed the geomagnetic induction  into the stationary Schr\"odinger equation form such that it becomes treatable by the standard exact Gel$'$fand-Levitan inversion procedure. 

In view of application of the \citet{gelfand1951} theory in its adapted \citet{weidelt1972} version  we adopt the Weidelt transformations
\begin{equation}
\xi :\ =\ k_0(0)=\sqrt{i\mu_0\omega\sigma_0}, \quad  \zeta(z,\xi):\ = \int_0^z \mathrm{d}y\sqrt{\Sigma(y,\xi)} 
\end{equation}
for the variables, and for the functions
\begin{eqnarray}
&& u(\zeta,\xi):\ =\ \Big[\Sigma(\zeta,\xi)\Big]^\frac{1}{4}\\[0.5ex]
&& f(\zeta,\xi):\ =\ u(\zeta,\xi){E_0(\zeta,\xi)} 
\end{eqnarray}
In these expressions the normalised frequency $\xi$ occurs as a parameter only. 

With $'$ now indicating differentiation with respect to $\zeta$ the conductivity-fluctuation-generalised induction equation transforms into the standard Sturm-Liouville (or stationary Schr\"odinger equation) form
\begin{equation}\label{eq-newind}
f''(\zeta,\xi)= \Big[\xi^2+V(\zeta,\xi)\Big]f(\zeta,\xi)
\end{equation}
This has been demonstrated by \citet{weidelt1972} by straightforward though somewhat tedious algebra (see Appendix 1). The ``equivalent potential'' function $V(\zeta,\xi)$ is defined solely and uniquely through the conductivity $\Sigma(\zeta,\xi)$ respectively $u(\zeta,\xi)$ as
\begin{equation}
V(\zeta,\xi) \equiv u''(\zeta,\xi)\big/ u(\zeta,\xi)
\end{equation}
It is this function which is considered unknown in the inversion problem. It is to be determined by inverting the induction equation for any  given frequency $\xi$ respectively $\omega$ or -- more generally -- any applied frequeny spectrum of the external geomagnetic induction field. On may, however, note that $\xi$ is just a parameter reserved for later use, not the primary important variable in the inversion problem. The relevant variable that enters the inversion is $\zeta$. 

Inverting the induction equation will yield the function $u(\zeta,\xi)$, from which one also obtains  $V(\zeta,\xi)$ on the way of differentiation. This finally provides the wanted dependence of $\Sigma(\zeta,\xi)$ on the depth $\zeta$ for given frequency $\xi$. Once this has been achieved, one encouters the difficult problem of interpreting the result in terms of the spectral density of the conductivity fluctuations. The advantage up to this point is that by relegating all the complicated distribution of the intrusions of conducting material to the fluctuation spectrum of  conductivity reduces the inverse geomagnetic induction problem to a strictly one-dimensional problem. Only the dependence on depth $\zeta$ enters. Any lateral dependence is reconstructed pointwise along a profile obtained at the Earth's surface. 

The important point to be made is that though $V(\zeta,\xi)$ is substantially more complicated than in the non-fluctuation models, there is no principal difference between those models and the present one what concerns the inversion theory. Once one has succeeded in determining $V(\zeta,\xi)$ from the data, the difficulty is shifted from the solution of the inversion problem to the interpretation of the obtained result, i.e. to the interpretation of $V(\zeta,\xi)$ in terms of a model of the subsurface structure of the conductivity distribution. It is the great achievement of \citet{weidelt1972} of having discovered the above transformations which have brought the induction equation into its general form Eq. (\ref{eq-newind}) making it treatable by the general inversion theory \citep{gelfand1951,marchenko2011} as had been developed for the wave equation and the inverse scattering problem. Since this theory is complete and has been shown to yield a unique solution, it is the appropriate way of solving the induction problem. \citet{parker1980} has discussed its pros and contras, its advantages and deficiencies.   

\citet{weidelt1972}, in following \citet{gelfand1951}, has shown that the solution of the inverse TE induction problem reduces to the solution of the linear Gel$'$fand-Levitan integral equation for some function $G(\zeta,\eta)$ (see Appendix 2)
\begin{equation}\label{eq-gleq}
G(\zeta,\eta)=K(\zeta +\eta) +\!\!\!\!\!\int\limits_{-\zeta}^{+\zeta} \mathrm{d}\tau G(\zeta,\tau)\Big\{K(\eta+\tau)+K(\eta-\tau)\Big\}  
\end{equation}
under the condition $|\eta| <\zeta$. It requires knowledge of the kernel function $K(\zeta)$, which itself is given by the measurements $c(\xi)$ at Earth's surface which is the inverse Laplace transform
\begin{equation}\label{eq-B}
K(\zeta)=\frac{1}{4\pi i}\int_{\epsilon-i\infty}^{\epsilon+\infty} \mathrm{d}\xi \Big[1-\xi c(\xi)\Big]\mathrm{e}^{\xi\zeta}
\end{equation}
Having determined $K(\zeta)$ from the observations contained in $c(\xi)$, the solution of the above Gel$'$fand-Levitan equation can be obtained by standard numerical methods iterating the integral equation, i.e. expanding its solution into a Neumann series. This procedure yields $G(\zeta,\eta)$. The solution of the inverse TE problem subsequently follows from
\begin{equation}
u(\zeta)= 1+\int_{-\zeta}^{+\zeta}\mathrm{d}\tau\ G(\zeta,\tau)
\end{equation}
It directly produces the vertical profile of the conductivity $\Sigma(\zeta,\xi)$. Straightforward differentiation with respect to $\zeta$ ultimately yields the wanted "equivalent potential" function $V(\zeta,\xi)$. 

Once the inverse TE problem has been solved, $u(\zeta,\xi)$ is locally given. This implies that $V(\zeta,\xi)$ and by it also the vertical profile $\Sigma(\zeta,\xi)$ are available at some site $x-x_0$ at Earth's surface. Pointwise construction of the vertical conductivity profile over an area of prospection on Earth's surface then, for given frequency $\xi$, allows obtaining the conductivity structure pointwise below that area of prospection. 

\section{Reconstruction of the fluctuations ?}
For practical purposes the above result suffices. It provides the vertical conductivity structure within a certain surface region. Its advantage is that it yields the corrections to a simple analytically solvable geophysically reasonable vertical conductivity-reference   model that had been initially assumed. 

This is what one would like to know from the practical point of view. One might, however, be interested in the reconstruction of the \emph{distribution} of conductivity itself, i.e. infer about the nature of the conductivity fluctuations. This is, unfortunately, not completely possible to achieve in detail for the reason that the information contained in the conductivity profile, though of practical use, is just  statistical. Inference about the nature of the fluctuations in real space requires not only the separation of the conductivity spectrum but also its re-transformation from wavelength/wavenumber space into real space. Below we show that this requires additional knowledge which might be available only in implicit form.
  
Applied to our case which involves fluctuations, the conductivity profile differs from the assumed large-scale reference-model profile $\sigma_0F_0(z)$. Since the relation between these two is linear, one can subtract the model profile. The deviation from it is the wanted contribution that is caused solely by the spectrum of conductivity fluctuations that is superimposed on the reference profile:
\begin{equation}
\Delta\Sigma=\Sigma - \sigma_0F_0 = \left[\frac{\xi^2}{2}\!\!\!\!\!\!\sum\limits_{n=1\atop k_n\neq +k_0}^m\!\!\!\!\!\!\frac{\left|\delta F_{k_n}\left(\sqrt{k_n^2-k_0^2}\right)\right|^2}{\sqrt{k_n^2-k_0^2}}\right]
\end{equation}
with left-hand side now known from the solution of the inverse problem. Here $k_0^2(\zeta,\xi)=\xi^2F_0(\zeta)$ contains the dependence of the fluctuations on $\zeta$ respectively $z$. One may remember that $\xi$ is complex. Similarly, the poles $k_n(\zeta,\xi)$, where the residua have been taken, are as well functions of $\zeta,\xi$.  So far they are unknown. Would the fluctuation spectrum be known, this would be the final result. This is, however, not the case. 

Determination of the conductivity fluctuations requires the inversion of the sum on the right and the determination of the various $k_n(\zeta)$ for a spectrum of frequencies $\xi$. 

There is only a finite number of poles and therefore only a finite number of wave numbers $k_n$. We also know that all $k_n=k_{n}^{r}+ik_{n}^i$ have positive imaginary parts $k_{n}^i>0$ only. Re-transformation from $k$-space into horizontal coordinates $x$ then yields
\begin{equation}\label{eq-deltasigma}
\Delta\Sigma(\zeta, x)= \frac{\xi^2}{4\pi}\sum\limits_{n=1}^m\frac{\mathrm{e}^{ik_n(x-x_0)}}{\sqrt{k_n^2-k_0^2}}\left|\delta F_{k_n}\left(\sqrt{k_n^2-k_0^2}\right)\right|^2
\end{equation}
in the one-dimensional case under consideration. The sum thus consists of a limited number $n\leq m$ of harmonic functions. For each harmonic we can write
\begin{equation}
\Delta\Sigma_n(\zeta,\xi) =  \frac{\xi^2}{4\pi}\frac{1}{\sqrt{k_n^2-k_0^2}}\left|\delta F_{k_n}\left(\sqrt{k_n^2-k_0^2}\right)\right|^2
\end{equation}
In this expression the dependence on depth $\zeta$ and frequency $\xi$ is explicitly contained in $k_0\sim\sqrt{\xi F_0(\zeta)}$. It is also  implicit to $k_n$ and thus not directly acessible without further assumptions. One realises, however, that the spectral density of conductivity fluctuations $|\delta F_{k_n}|^2$ could in principle be determined if only the dependence of $k_n$ on depth and frequency would be known. 
 
Determination is possible of the horizontal wavelengths $\lambda_n=2\pi/k_{n}^r$ of the contributing conductivity fluctuations at the surface of Earth at $z=0$ for each applied real frequency $\xi$ as well as their surface amplitudes. This requires the measurement of $c(\xi)$ as function of $x-x_0$ in an area on the surface along the direction of dominant variation.  (A similar procedure would be possible also for two-dimensional dependencies.) 

It is reasonable to assume that the real parts  $k_{n}^r$ of the wavenumbers $k_n$ do not (or only weakly) depend on depth $\zeta$. If they are independent on $z$ then the variation of the amplitudes of the fluctuation spectrum is included in the dependence of $k_0(z)$ which is given by $F_0(z)$. 
In this approximation, determination of $k_{n}^r(z=0)$ fixes the real part of the wavenumbers of the fluctuations over a certain range of depths. 

The imaginary part $k_{n}^i(\xi,\zeta=0)$ for given $n$ and frequency $\xi$ on the Earth's surface can be determined from the behaviour of \begin{equation}
\Delta\Sigma(x-x_0,\zeta=0)\sim \exp[-k_{n}^i(x-x_0)] 
\end{equation}
along the profile $x-x_0$. For $k_{n}^i>0$ the amplitude $\Delta\Sigma_n$ \emph{increases} as long as $x<x_0$. It reaches maximum at $x=x_0$ and decays for $x>x_0$. Hence, $x_0(n,\xi)$ is identified as the wavelength and frequency dependent location of the maximum response in $\Delta\Sigma$ at the surface. Ideally this behaviour is exponential. It should be treated by considering the logarithmic profile. Its slope provides the imaginary part $k_{n}^i(\zeta=0,\xi)$, however only at the surface for any given frequency $\xi$. 

In contrast to the wavelength and real part $k_{n}^r$, the imaginary part $k_{n}^i(\zeta,\xi)$ is, however, a function of depth $\zeta$ and frequency $\xi$ because it contains the spatial damping/penetration effect of the conducting layers. The determination of its dependence on $\zeta$ poses a major problem which cannot be solved easily. For this reason the explicit determination of the conductivity fluctuations in real space is practically impossible from the measurement of the response function at Earth's surface. One has to stay with the information contained in $\Delta\Sigma(\zeta,\xi)$. 

There is no obvious way of how $k_{n}^i(\zeta)$ could be inferred \emph{precisely}. This inability inhibits the precise functional reconstruction of the conductivity fluctuations, i.e. it inhibits the precise knowledge about the physical nature of the fluctuations. The solution of the inverse problem only provides the contribution of the fluctuation spectrum to the conductivity, it does not say for whatever \emph{geophysical} reason the conductivity structure has the partucular spatial distribution obtained from the solution of the fluctuation-generalised Helmholtz equation and the related inverse problem. The re-transformation of the fluctuation spectrum into real space and inference about the fluctuations themselves in inhibited.  

In order to proceed one must introduce further assumptions. One may, for instance, assume that $k_n^i(\zeta,\xi)$ at a given surface point $x-x_0$ is related to the penetration depth of the induction field of frequency $\omega\sim \Im(\xi)$. This assumption is reasonable in the sense of an approximation. Identifying $k_n^i$ with the penetration depth one then has 
\begin{equation}\label{eq-kni}
k_n^i(\xi,\zeta)=2\pi\, \Re\left[\sqrt{\xi\Sigma(\xi,\zeta)}\right] >0
\end{equation}
positive and real, with $\Sigma(\zeta)$ the total conductivity below the prospection point $x-x_0$. This identification is suggested by the fact that the applied electromagnetic induction field can penetrate only over a finite distance into the conducting Earth. Adopting this assumption we have for the argument of the root in the above expressions
\begin{equation}
k_n^2-k_0^2= \Big({k_n^r}^2-{k_n^i}^2\Big)-i\Big(\rho_0^2-2k_n^rk_n^i\Big)
\end{equation}
Multiplying Eq. (\ref{eq-deltasigma}) with the root yields an implicit representation of the spectral density for each harmonic $n$ as 
\begin{equation}
\frac{1}{4\pi}\Big|\delta F_{n}(\zeta,\xi)\Big|^2 =\Re\,\bigg[\sqrt{k_n^2-k_0^2}\frac{\Delta\Sigma_n}{\xi^2}\bigg]
\end{equation}
The right-hand side of this expression is known because the real part of $k_n$ is given from the inspection of the horizontal variation of the Earth's response, and the imaginary part of $k_n$ is determined by its relation to the penetration depth which is a function of the vertical conductivity profile and frequency as given in Eq. (\ref{eq-kni}). With known complex conductivity profile $\Delta\Sigma=\Sigma -\sigma_0F_0$, as obtained from the solution of the inverse problem for a given reference profile $\sigma_0F_0$, this yields the spectral density of the fluctuations as function of depth and frequency, i.e. the spectral power in each harmonic $n$, at a given point $x-x_0$ along an $x$-profile on the surface. This is the maximum information that can be obtained about the fluctuations. It is subject to further geophysical interpretation.

From the point of view of the TE problem in Magneto-Tellurics and geomagnetic induction the impossibility of reconstruction of the fluctuations themselves from the spectral density is not a drawback. Determination of the average conductivity profile suffices for all practical purposes. The additional knowledge of the depth dependence of the spectral density just provides information about the regions where and where not the conductivity fluctuations contribute.

\section{Conclusions}
The present note extends the TE problem of geomagnetic induction to the inclusion of a spectrum of spatial conductivity fluctuations in the underground. 

This is a fundamentally different approach to the induction problem. Its assumption is that given an average reference profile of the vertical conductivity distribution $\sigma_oF_0(z)$, which can be chosen as an reasonable initial, analytically treatable model of the dependence of the conductivity on depth, a superimposed statistical distribution of spatially unresolved and \emph{principally unresolvable} differently conducting regions of various scales is shown to contribute to the average conductivity profile. This contribution appears in terms of the spectral density of the distribution of fluctuations. 

In an approach which prescribes the distribution, the forward problem should be solved. Such an approach, however, corresponds to the ordinary forward problem and thus does not provide any advantage over the usual theory. On the other hand, leaving the fluctuations undetermined, i.e. treating them as an unknown distribution of conductivity fluctuations, solution of the inverse TE problem by application of any of the inverse methods that have been developed in the past, yields an explicit -- in the majority of cases numerially given -- expression for the fluctuation-caused conductivity. 

This we have demonstrated by reference to Weidelt's (1972) treatment of the inverse geomagnetic induction problem. But any of the known approaches to solve the inverse problem would do as well as long as it provides a set of local conductivity profiles over a surface area. The solution of the inverse problem then yields a local depth profile of the fluctuation-mediated conductivity in the subground which is superimposed on the chosen reference model. Within the analytical method of \citet{weidelt1972} these expressions have been given by us. Clearly, by modifying the initial reference profile, the profiles could be optimised. Moreover, since the fluctuation model includes the lateral dependence of the conductivity spectrum, exploration of the profile pointwise over a surface area for different frequencies $\omega$ of the induction field provides information about the lateral variation of the fluctuation-modified profile. It also provides information about the spectrum of wavelengths of the conductivity fluctuations.   

\vspace{1.5cm}
\section{Appendix 1: Derivation of Equation (\ref{eq-newind})}
Here, for completeness and since the algebra is nontrivial, we provide the transformation of the induction equation 
\begin{equation}
A_{xx}=\xi^2\Sigma(x) A\quad\mathrm{with}\quad A_x=\mathrm{d}A/\mathrm{d}x
\end{equation}
into the Schr\"odinger-like form
\begin{equation}
f''(\zeta)=\Big[\xi^2+V(\zeta)\Big]f(\zeta),\ \ \ \zeta=\int_0^x\mathrm{d}t\sqrt{\Sigma(t)}
\end{equation}
Define $u(\zeta)\ :\ =\sqrt[4]{\Sigma(\zeta)}$ and $f(\zeta)=u(\zeta)A(\zeta)$. Then one has $\mathrm{d}\zeta/\mathrm{d}x=u^2(\zeta)$.  Divide the above original induction equation by $\Sigma=u^4$ and use\ $' : \ =\mathrm{d}/\mathrm{d}\zeta$ to obtain
\begin{eqnarray}
f'&=&u'A+uA' \nonumber\\
f''&=& u''A+2u'A'+uA''\nonumber
\end{eqnarray}
Now, $A'=A_x/u^2$. Thus
\begin{equation}
f''=u''A + \Big\{2(u'/u^2)A_x +uA''\Big\} 
\end{equation}
One must now express $uA''$ in order to recover $A_{xx}$. This is done by calculation as follows:
\begin{equation}
A'' =\bigg(\frac{A_x}{u^2}\bigg)'=\frac{A_{xx}}{u^4}-\frac{2u'}{u^3}A_x
\end{equation}
Multiply by $u$ and use in the first equation and in the curly brackets of the last above expressions for $A_{xx}$ and  $f''$, rearrange and define $V(\zeta)=u''(\zeta)/u(\zeta)$. This produces the wanted Schr\"odinger form of the original induction equation. It is clear that this form cannot be trivially obtained. 

In order to finally, after solving the inverse problem, recover the spatial coordinate $x$ one has to perform the integral
\begin{equation}
x(\zeta) = \int_0^\zeta\mathrm{d}t/u^2(t)
\end{equation}
This can be done because the solution of the Gel$'$fand-Levitan inverse problem produces $u(\zeta)$ from the observational data.

\newpage
\section{Appendix 2: The Gel$'$fand-Levitan equation}
In this Appendix we, for completeness, give a brief derivation of the Gel$'$fand-Levitan equation in the form applicable to the TE geomagetic induction problem as used in this paper. The reason for repeating it here is that the original publication of \citet{weidelt1972} is hardly available anymore. It therefore makes sense to at least cursorily repeat the steps on which the particular form of the Gel$'$fand-Levitan equation is obtained. This form differs from the original one.  \citet{weidelt1972} adapted it to the needs of the geophysical application while its original version was formulated for the scattering problem in quantum electrodynamics.

The second order Schr\"odinger differential equation (\ref{eq-newind}) has two solutions $f_\pm(\zeta,\xi)$ which satisfy the boundary conditions $f_\pm(0,\xi)=1, f'_\pm(0,\xi)=u'(0)\pm\xi$. These solutions are the \emph{Jost-solutions} \citep[cf., e.g.,][]{koelink2008}
\begin{equation}
f_\pm(\zeta,\xi)=\mathrm{e}^{\pm\xi\zeta}+\int_{-\zeta}^{+\zeta}\mathrm{d}t\ G(\zeta,t)\ \mathrm{e}^{\pm\xi t}
\end{equation}
They are obtained replacing the second-order differential Sturm-Liouville (Schr\"odinger) equation by  an inhomogeneous differential equations with inhomogeneity $g=Vf$, using the method of variation of constants. This solution is also called the \emph{Schr\"odinger integral equation}. The function $G(\zeta,t)$ is subject to the conditions $G(0,0)=\frac{1}{2}u'(0)=G(\zeta,-\zeta)$. It satisfies the differential equation $G''-G_{tt}=V(\zeta)G$ and thus is related to the unknown potential by
\begin{equation}
G(\zeta,\zeta)-\frac{1}{2}u'(0)=\frac{1}{2}\int_0^\zeta\mathrm{d}t\ V(t)
\end{equation}
The relation between $G$ and $u$ is that
\begin{equation}
u(\zeta)=\lim\limits_{\xi\to0} f_\pm(\zeta,\xi)= 1+\int_{-\zeta}^{+\zeta}\mathrm{d}t\ G(\zeta,t)
\end{equation}
The full solution of the Schr\"odinger equation is a linear combination of the two solutions $f_\pm$. Defining $b(\xi)=\frac{1}{2}\big[1-\xi c(\xi)\big]$ and using the boundary conditions this can be expressed in terms of the observational function $c(\xi)$ as follows:
\begin{equation}
\big[1-2b(\xi)\big]f(\xi,\zeta)=f_-(\xi,\zeta)-b(\xi)\big[f_+(\xi,\zeta)+f_-(\xi,\zeta)\big]
\end{equation}\newpage
If we use the Jost solution, then this yields explicitly
\begin{eqnarray}
\big[1&-&2b(\xi)\big]f(\xi,\zeta)-\mathrm{e}^{-\xi\zeta}\ =\ \int_{-\zeta}^\zeta\mathrm{d}t\ G(\zeta,t)\ -\cr 
&-&2b(\xi)\left[\cosh(\xi\zeta) +\int_{-\zeta}^\zeta\mathrm{d}t\ G(\zeta,t)\cosh{\xi t}\right]
\end{eqnarray}
One now multiplies by $\exp(\xi\eta)/2\pi i$ and integrates in the complex $\xi$-plane with respect to $\xi = \epsilon > 0$ which leads to four integrals (inverse Laplace transforms) in the limit $\xi\to\infty$. The left-hand side taken over the right half-plane gives simply zero, because there are no poles and the solution must for physical reasons be assumed analytical for $|\eta|<\xi$. The second integral which is over the first term on the right, simply compensates for the integral such that it yields $G(\zeta,\eta)$. The third integral, taken of the second term, contains the pure data and thus just yields a function $-K(\zeta+\eta)$ where
\begin{equation}
K(\zeta)=\frac{1}{2\pi i}\int\limits_{\epsilon-i\infty}^{\epsilon+i\infty}\mathrm{d}\xi b(\xi) =\frac{1}{4\pi i}\int\limits_{\epsilon-i\infty}^{\epsilon+\infty} \mathrm{d}\xi \Big[1-\xi c(\xi)\Big]\mathrm{e}^{\xi\zeta}
\end{equation}
is the inverse Laplace transform of the data function as given in Eq. (\ref{eq-B}). This is assumed known but provides the main difficulty in the inversion procedure because the calculation of the inverse Laplace transform using measurements is a nontrivial problem \citep{parker1980}. Finally, the integral taken over the last term in the above expression is a Laplace convolution integral between the functions $G$ and $K$, the latter containing the data and forming the kernel, the former being the kernel in the above Schr\"odinger integral equation which is assumed known but must be obtained from solving the integral equation. Thus it is represented as
\begin{equation}
-\int_{-\zeta}^{+\zeta}\mathrm{d}t\ G(\zeta,t)\big[K(\eta+t)+K(\eta-t)\big]
\end{equation}
Combining all the integrals ultimately yields the Gel$'$fand-Levitan integral equation (\ref{eq-gleq}) for the function $G(\zeta,\eta)$ in dependence on the kernel function $K(\zeta,\eta)$ which is a functional of the given observational data.


\newpage

\begin{thebibliography}{99}

\bibitem[Alumbaugh et al.(1996)]{alumbough1996} Alumbaugh DL, Newman GA, Prevost L and Shadid JN (1996) Three-dimensional wideband electromagnetic modeling on massively parallel computers, Radio Sci 31:1-23, doi: 10.1029/95RS02815.

\bibitem[Avdeev(2005)]{avdeev2005} Avdeev BD (2005) Three-dimensional electromagnetic modelling and inversion from theory to application, Surv Geophys 26:767-799, doi: 10.1007/s10712-005-1836-x.



\bibitem[Backus and Gilbert(1967)]{backus1967} Backus GE and Gilbert JF (1967) Numerical applications of a formalism for geophysical inverse problems, Geophys J RAS 13:247-276, doi: 10.1111/j.1365-246X.1967.tb02159.x.

\bibitem[Bailey(1970)]{bailey1970} Bailey RC (1970) Inversion of the geomagnetic induction problem. Proc Roy Soc London Ser A 315:185-194, doi: 10.1098/rspa.1970.0036.

\bibitem[Egbert and Kelbert(2012)]{egbert2012} Egbert GD and  Kelbert A (2012) Computational recipes for electromagnetic inverse problems, Geophys J Int 189:167-251, doi: 10.1111/j.1365-246X.2011.05347.x.



\bibitem[Gel$'$fand and Levitan(1951)]{gelfand1951} Gel$'$fand IM and Levitan BM (1951) On the determination of a differential equation from its spectral function, Izv Akad Nauk SSSR, mat ser 15, 309-360 [Engl. Transl. in: Amer Math Soc, Transl Ser 2, 1, 253-304, 1955] 

\bibitem[Gubbins(2004)]{gubbins2004} Gibbins D (2004) Time Series Analysis and Inverse Theory for Geophysicists, Cambridge University Press, Cambridge UK.

\bibitem[Jackson(1972)]{jackson1970} Jackson DD (1972) Interpretation of inaccurate, insufficient and inconsistent data, Geophys J RAS 28:97-109, doi: 10.1111/j.1365-246X.1972.tb06115.x.

\bibitem[Johnson and Smylie(1971)]{johnson1971} Johnson IM and Smylie DE (1971) An inverse theory for the calculation of the electrical conductivity of the lower mantle. Geophys J RAS 22:41-53, doi: 10.1111/j.1365-246X.1971.tb03582.x.

\bibitem[Kelbert et al(2014)]{kelbert2014} Kelbert A, Kuvshinov A, Velimsky J, Koyama T, Ribaudo J, Sun J, Martinec Z and Weiss CJ (2014) Global 3-D electromagnetic forward modelling: a benchmark study, Geophys J 197:785-814, doi: 10.1093/gji/ggu028.


\bibitem[Koelink(2008)]{koelink2008} Koelink E (2008) Scattering theory. Lecture Notes, Spring 2008, Radboud U Nijmegen, NL

\bibitem[Marchenko(2011)]{marchenko2011} Marchenko VA (2011) Sturm-Liouville operators and applications, Amer Math Soc, Chelsea Publ, Providence RI, USA.

\bibitem[Meqbel et al(2014)]{meqbel2014} Meqbel NM, Egbert GD, Wannamaker PE, Kelbert A and Schultz A (201) Deep electrical resistivity structure of the Pacific Northwestern U. S. derived from 3-D inversion of US Array Magnetotelluric data, Earth Planet Sci Lett 402:290-304, doi: 10.1016/j.epsl.2013.12.026.


\bibitem[Parker(1971)]{parker1970} Parker RL (1971) The inverse problem of electrical conductivity in the mantle, Geophys J RAS 22:121-138, doi: 10.1111/j.1365-246X.1971.tb03587.x

\bibitem[Parker(1980)]{parker1980} Parker RL (1980) The inverse problem of electromagnetic induction -- Existence and construction of solutions based on incomplete data, J Geophys Res 85:4421-4428, doi:10.1029/JB085iB08p04421.

\bibitem[Parker(1994)]{parker1994} Parker RL (1994) Geophysical Inverse Theory, Princeton University Press, Princeton NJ, USA.

\bibitem[Beusekom et al.(2010)]{beusekom2010} Van Beusekom AE, Parker RL, Bank RE, Gill PE, Constable S (2010) The 2-D magnetotelluric inverse problem solved with optimization, Geophys J Int 184:639-650, doi: 10.1111/j.1365-246X.2010.04895.x

\bibitem[Weidelt(1972)]{weidelt1972} Weidelt P (1972) The inverse problem of geomagnetic induction, Z Geophys 38:257-289.

\bibitem[Weidelt(2005)]{weidelt2005} Weidelt P (2005) The relationship between the spectral function and the underlying conductivity structure in 1-D magnetotellurics, Geophys J Int 161:566-590, doi:10.1111/j.1365-246X.2005.02625.x

\bibitem[Yang et al(2015)]{yang2015} Yang B, Egbert GD, Kelbert A and Meqbel NM (2015) Three-dimensional electrical resistivity of the north-central USA from EarthScope long period magnetotelluric data, Earth Planet Sci Lett 422:87-93, doi: 	
10.1016/j.epsl.2015.04.006.

\end{thebibliography}
\end{document}